\documentclass[twocolumn]{aastex631}

\newcommand{\chandra}{Chandra}

\newcommand{\xmm}{XMM-Newton}

\newcommand{\nustar}{NuSTAR}

\newcommand\zwcl{ZWCL 1856.8}
\usepackage{multirow}
\usepackage{wrapfig} 
\usepackage{lineno}
\revised{March 12, 2026}

\shorttitle{X-ray evidence from NuSTAR for a Mach 3 shock in Merging Galaxy Cluster \zwcl}
\shortauthors{T\"{u}mer et al.}

\begin{document}

\title{X-ray evidence from NuSTAR for a Mach 3 shock in Merging Galaxy Cluster \zwcl}

\correspondingauthor{Ay\c{s}eg\"{u}l T\"{u}mer}
\email{aysegultumer@gmail.com}

\author[0000-0002-3132-8776]{Ay\c{s}eg\"{u}l T\"{u}mer}
\affiliation{Center for Space Science and Technology, University of Maryland, Baltimore County (UMBC), Baltimore, MD 21250, USA}
\affiliation{NASA / Goddard Space Flight Center, Greenbelt, MD 20771, USA}
\affiliation{Center for Research and Exploration in Space Science and Technology, NASA / GSFC (CRESST II), Greenbelt, MD 20771, USA}  

\author[0000-0001-9389-6050]{Christian T. Norseth}
\affiliation{Department of Physics \& Astronomy, The University of Utah, 115 South 1400 East, Salt Lake City, UT 84112, USA}

\author[0000-0001-9110-2245]{Daniel R. Wik}
\affiliation{Department of Physics \& Astronomy, The University of Utah, 115 South 1400 East, Salt Lake City, UT 84112, USA}

\begin{abstract}
We present spectral analysis results of deeper (270~ks) NuSTAR observations of the merging galaxy cluster system, ZWCL1856.8+6616, at redshift $z = 0.304$, following a pilot study using shallower (30~ks) NuSTAR data \citep{tumer24}. The cluster hosts a double radio relic, pointing to a similar mass head-on collision at/near the plane of sky. We aim to find the relation between radio and X-ray shock features. Using data from both focal plane modules of NuSTAR, we study the temperature structure across the field of view and report on the X-ray detected shock strength at the relic sites. We generate nominal and cross-ARFs with \textsc{nucrossarf} to disentangle photon cross-contamination within regions of interest due to the moderate point spread function of NuSTAR. Here we report one of the strongest X-ray detected shocks in a galaxy cluster merger with $\mathcal{M}$=$3.90^{+1.64}_{-0.85}$ at the Northern relic site, that is unprecedentedly larger than the radio counterpart; $\mathcal{M}$=2.5$\pm$0.2 \citep{jones21}, and we report Southern shock strength as $\mathcal{M}$=$2.36^{+0.58}_{-0.46}$. We argue that since the Northern relic (or radio shock), is confined in a very small region in the sky, particle acceleration is more efficient and is likely to grow in the post-shock regions. In addition, we search for inverse Compton (IC) emission at the radio relic sites; however, an IC component was not detected.
\end{abstract}

\keywords{X-rays: galaxies: clusters --- galaxies: clusters: individual (\zwcl), intracluster medium --- radiation mechanisms: non-thermal, thermal --- shock waves}

\section{Introduction} \label{sec:intro}
Galaxy clusters, the most massive gravitationally bound structures in the Universe, evolve through a sequence of hierarchical mergers and accretion events that release energies up to $\sim 10^{64}$--$10^{65}$ erg over gigayear timescales \citep{Sarazin2002, Kravtsov2012}. These merger events drive large-scale hydrodynamic disturbances in the intracluster medium (ICM), giving rise to phenomena such as shocks, cold fronts, and turbulence \citep[e.g.,][]{Markevitch2007}. Among these, shock fronts are of particular astrophysical interest because they directly probe the conversion of bulk kinetic energy into thermal energy, the acceleration of cosmic rays, and the amplification of magnetic fields \citep{Brunetti2014, Vazza2012}.

Radio relics are large-scale, diffuse synchrotron sources observed predominantly in cluster outskirts. They trace relativistic electrons accelerated at shock fronts in the presence of magnetic fields, producing steep-spectrum radio emission. Their morphology, polarization properties, and spectral steepening across the relic width provide valuable insights into the geometry and aging of the electron population. If an X-ray-identified shock coincides spatially with a radio relic, it strongly supports the scenario in which the relic was formed by a merger-driven shock, thereby validating models of diffusive shock acceleration (DSA) \citep{Brunetti2014,Botteon2016,Hoang2018}.

In the X-ray regime, shocks manifest as surface brightness and temperature discontinuities. The Rankine-Hugoniot conditions provide a means to quantify the shock strength through the Mach number $\mathcal{M}$, estimated from the temperature jump across the front as:
\begin{equation}
\frac{T_2}{T_1} = \frac{5\mathcal{M}^4 + 14\mathcal{M}^2 - 3}{16\mathcal{M}^2},
\label{eq:mach_temp}
\end{equation}
where $T_2$ and $T_1$ denote the post- and pre-shock temperatures, respectively. Shock speeds in clusters typically range from 1000--3000~km~s$^{-1}$, with resulting Mach numbers rarely exceeding 3 due to the high pre-shock sound speeds in the hot ICM \citep{Ryu2003, Hong2015}.
Double radio relic systems are particularly valuable because they provide a near-symmetric probe of merger kinematics. These systems are thought to result from near head-on collisions between two comparable massive subclusters in the plane of the sky \citep{bonafede12,deGasperin14}. They allow constraints to be put on the merger axis, impact parameter, and dynamical age of the system. While the relics in such systems are often located at roughly equal distances from the cluster center, their morphological symmetry is not guaranteed. Disparities in extent and surface brightness can arise due to differences in the ambient medium, shock strength, or projection effects \citep{Kang2015,weeren19}. A smaller spatial extent may indicate an aging or less energetic shock, or possibly a relic that has encountered a denser medium which has decelerated the shock front \citep{shimwell15, DiGennaro2021}.

ZWCL1856.8+6616 \citep[or PSZ1 G096.89+24.17,][hereafter ZWCL1856.8]{planck16}, at $z = 0.304$, represents one of the few known examples of a double radio relic system \citep{deGasperin14,finner21}. The presence of two prominent relics located symmetrically about the cluster core strongly suggests a major merger viewed close to the plane of the sky. Previous \textit{Chandra} and \textit{XMM-Newton} observations indicated complex temperature structures suggestive of shock activity, but lacked the depth required to definitively identify shock fronts or constrain IC emission \citep{finner21, Jones2021}. In this work, we present new, deeper observations with \textit{NuSTAR}, enabling a more sensitive analysis of the thermodynamic structure of the ICM and a detailed investigation of potential shock-relic associations in this exceptional cluster merger system. We refer to our pilot study \citep{tumer24} for an extended introduction to the system.

Throughout this paper, we assume the $\Lambda$CDM cosmology with {\it H$_{0}$} = 70 km s$^{-1}$ Mpc$^{-1}$, $\Omega_{M} =0.27$, $\Omega_{\Lambda} =  0.73$. According to these assumptions, at the cluster redshift, a projected intracluster distance of 100 kpc corresponds to an angular separation of $\sim$22\arcsec. All uncertainties are quoted 1$\sigma$ unless otherwise stated.

\section{Observations and data reduction} \label{sec:reduction}

In this work, we used \nustar~\citep[Nuclear Spectroscopic Telescope Array;][]{harrison13} observations of \zwcl, with observation ID 70901003002. This observation took place using both focal plane modules (i.e., FPMA and FPMB) during 22--27 October 2023 ($\sim$ 290 ks). \nustar~also observed the same source during 25--26 July 2022 ($\sim$30 ks) with OsbID 70801003002 as part of a pilot study to test NuSTAR's capability to constrain observables of multi-phase emission from regions away from the optical axis, in order to request sufficient and realistic exposure times for the following NuSTAR guest observer program. This was the first NuSTAR pointing of a merging galaxy cluster which encompassed a double radio relic within the field of view (FOV). The results of the pilot study are published in 2024 \citep{tumer24}. After pre-liminary assessment, we decided to exclude the short observation from the data analysis in this work given the low S/N of the aforementioned observation as well as extremely complicated multi region spectral fitting resulting in expensive computational time (folding models through 2$\times$ the number of Ancillary Response Files (ARFs)).

The data were filtered using the standard processing pipeline with HEASoft (v.~6.32) and \textsc{nustardas} (v.~2.1.2) tools along with NuSTAR CALDB version 20231003. Stage 1 and 2 of the \textsc{nustardas} processing pipeline script \textsc{nupipeline} was used to clean the event files. The screening criteria follows the same procedure described in detail in \citet{tumer24}. 
After filtering for periods of high background counts, the cleaned data had a reduced exposure time of 264 ks.

\begin{figure*}
\centering
\includegraphics[width=180mm]{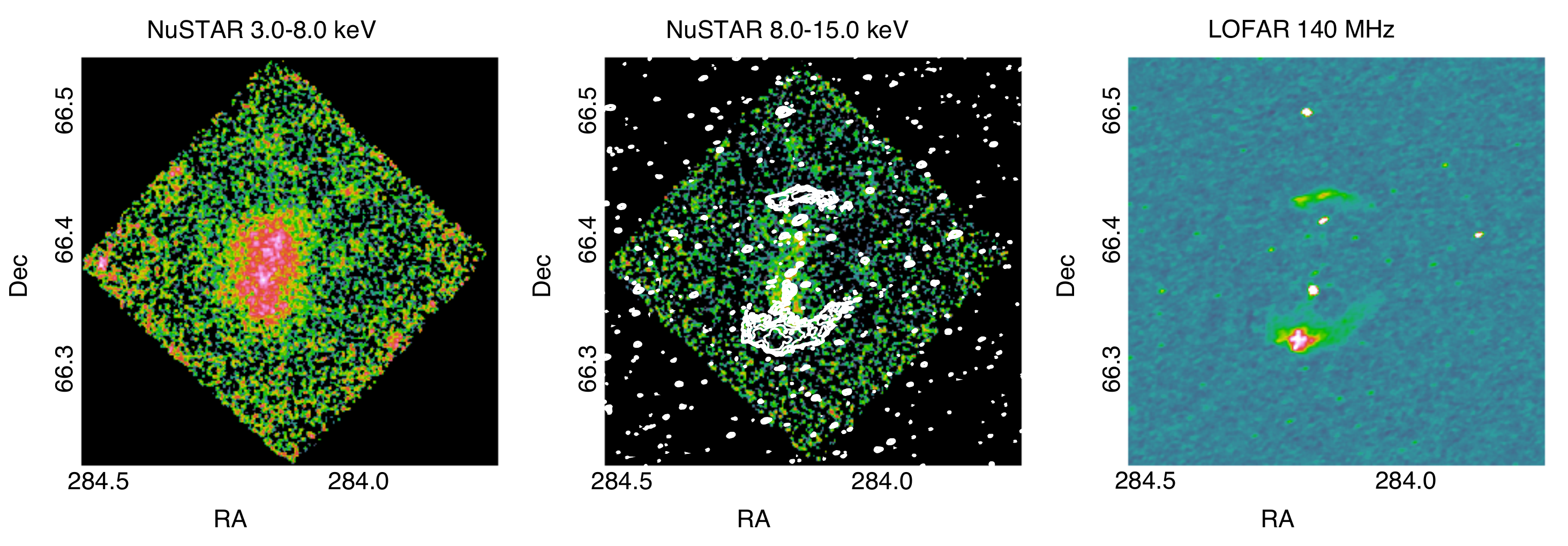}
\caption{Background subtracted, exposure corrected, \nustar~images of \zwcl~smoothed with a Gaussian of r=5 pixels (1 pixel subtending 12$\arcsec$.3) in the soft ({\it left}) and hard bands ({\it middle}). {\it Right:} and LOFAR image at 140 MHz (resolution: 13$\arcsec$ $\times$ 7$\arcsec$, RMS noise: 127 $\mu$Jy/beam).
\label{fig:lofarnustarphoton}}
\end{figure*}

We used \textsc{nuskybgd} \citep{wik14} for the assessment of background. The global background model was then used to create background images as well as background spectra for regions of interest (ROIs). We describe the background treatment in the Appendix section of this work. Background subtracted, exposure corrected images in the 3.0~--~8.0 keV and 8.0~--~15.0 keV bands and are presented in the left and middle panels of Figure~\ref{fig:lofarnustarphoton}.

\section{Data Analysis and Results} \label{sec:analysis}
In this paper, we aim to study the correlation of the ICM features of \zwcl\ revealed by the radio and the X-ray data. The outer edges of the radio relics detected by WSRT \citep{deGasperin14} and LOFAR \citep{jones21} are expected to coincide with X-ray shocks that can be revealed as temperature and density jumps across the shock front. Due to \nustar's moderate point spread function (PSF), our analysis includes an assessment of the contamination from each region of interest into the others in order to put robust constraints on the model parameters in regions with low S/N. Since good constraints on the temperatures are required to assess the Mach number related to the shocks, this contamination or ``cross-talk" becomes important when analyzing regions with low S/N  such as cluster outskirts where unshocked plasma resides.

In the next two subsections, we describe the selection of the regions of interest along with the selection of the best fit models.
We then present how the cross-talk analysis is conducted.

\subsection{Setting The Stage For \textsc{nucrossarf}: Selection of Regions of Interest and Designation of Initial Parameters }\label{sec:specanalysis}

\subsubsection{Region Selection}
\begin{figure}
\centering
\includegraphics[width=80mm]{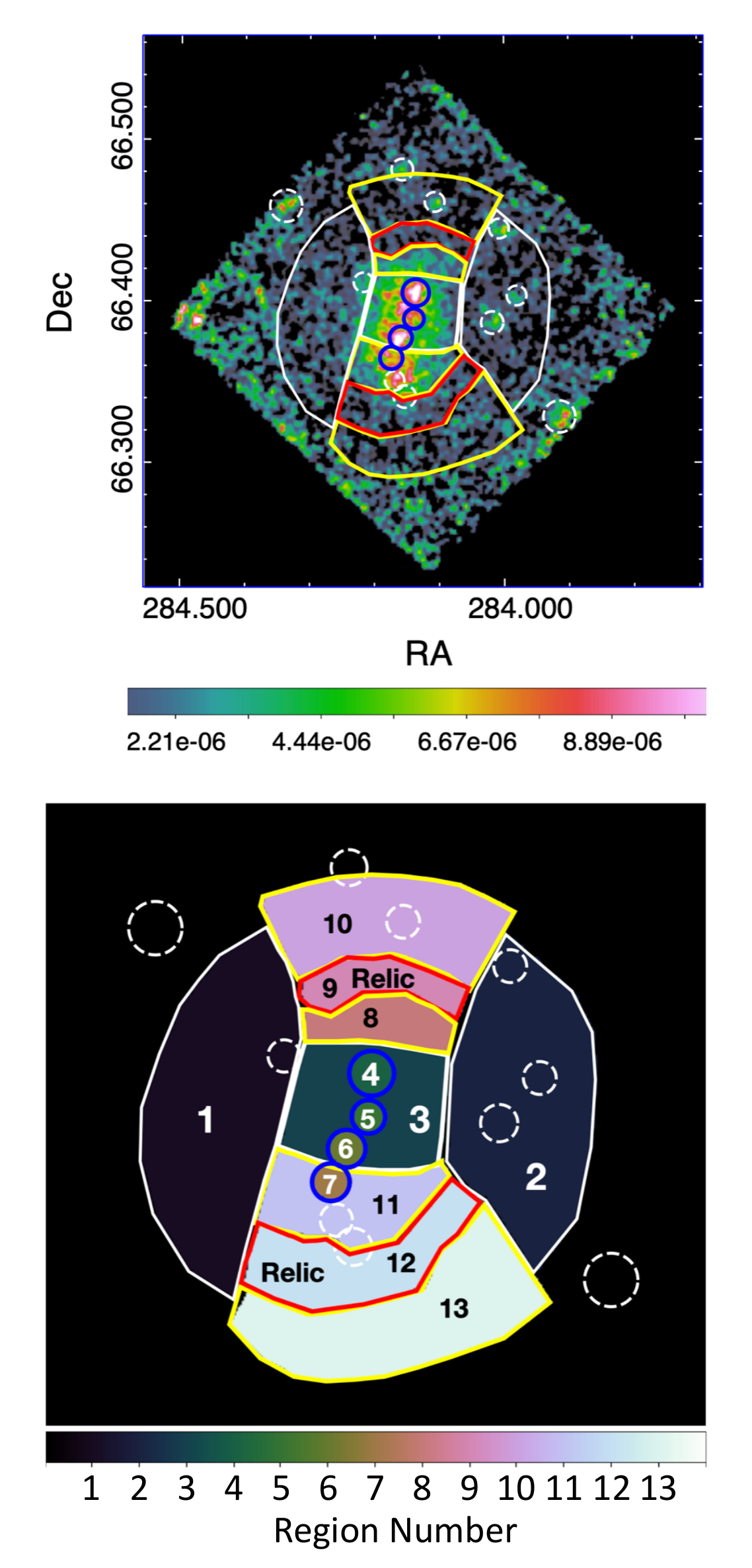}
\caption{\textit{Top Panel:} Selected regions overlaid on background subtracted, exposure corrected images {\it Bottom Panel}: Region numbers.
\label{fig:ROI}}
\end{figure}

We selected 13 regions of interest that show distinct emission features based on optical, radio and X-ray data, overlaid on the NuSTAR count rate image in Fig.~\ref{fig:ROI}. We defined two regions corresponding to the locations of the radio relics seen by LOFAR in the North (Region 9, NR) and South (Region 12, SR) shown in the middle and right panels of Fig~\ref{fig:lofarnustarphoton}, in order to assess non-thermal contributions to the observed emission. 

Additionally, we chose two regions at the inner and outer part of each relic region to assess the temperature around the radio shocks: Region 10: North of the NR covering pre-shock gas, Region 8: South of the NR covering post-shock gas, Region 11; North of the SR (post-shock), and Region 13; South of the SR (pre-shock). Although the consensus dictates that Radio relic regions are part of shocked plasma, we separate them from Regions 10 and 13 for the purpose of searching for non-thermal emission. Furthermore, because of NuSTAR's moderate angular resolution, the exact location of the possible surface brightness jump is unknown. Nevertheless, we present an analysis in Section~\ref{sec:cross-talk} where post-shock regions are considered as the combination of Region 8 + Region 9 for the North, and Region 11 + Region 12 for the South.

The pre-shock region outer edge selection was based on the decision to include as much emission as possible without including the pixels near the edge of the FOV that contain bright pixels. The inner edges of the post-shock regions were determined by the Brightest Cluster Galaxies (BCG) positions. 

We selected a central region that encompasses the brightest X-ray emission (Region 3, Center), but excluded Regions 4 and 6 that encompass the two BCGs as well as Regions 5 and 7, which are Chandra and NuSTAR resolved point sources \citep{tumer24}. 

Two additional regions were defined to encompass the remaining emission of the cluster in the East (Region 1) and the West (Region 2).

Other point sources (most likely AGN) are detected within these regions and are visible in the top panel of Figure~\ref{fig:ROI}. The dashed line circles in Fig.~\ref{fig:ROI} indicate their positions. 
The rationality of the numbering scheme lies in the intention of a facilitated reading of the analysis result tables, which will become apparent in the next sections.

\subsubsection{Assessment of Initial Fit Parameter Values}

\begin{figure}
\centering
\includegraphics[width=96mm]{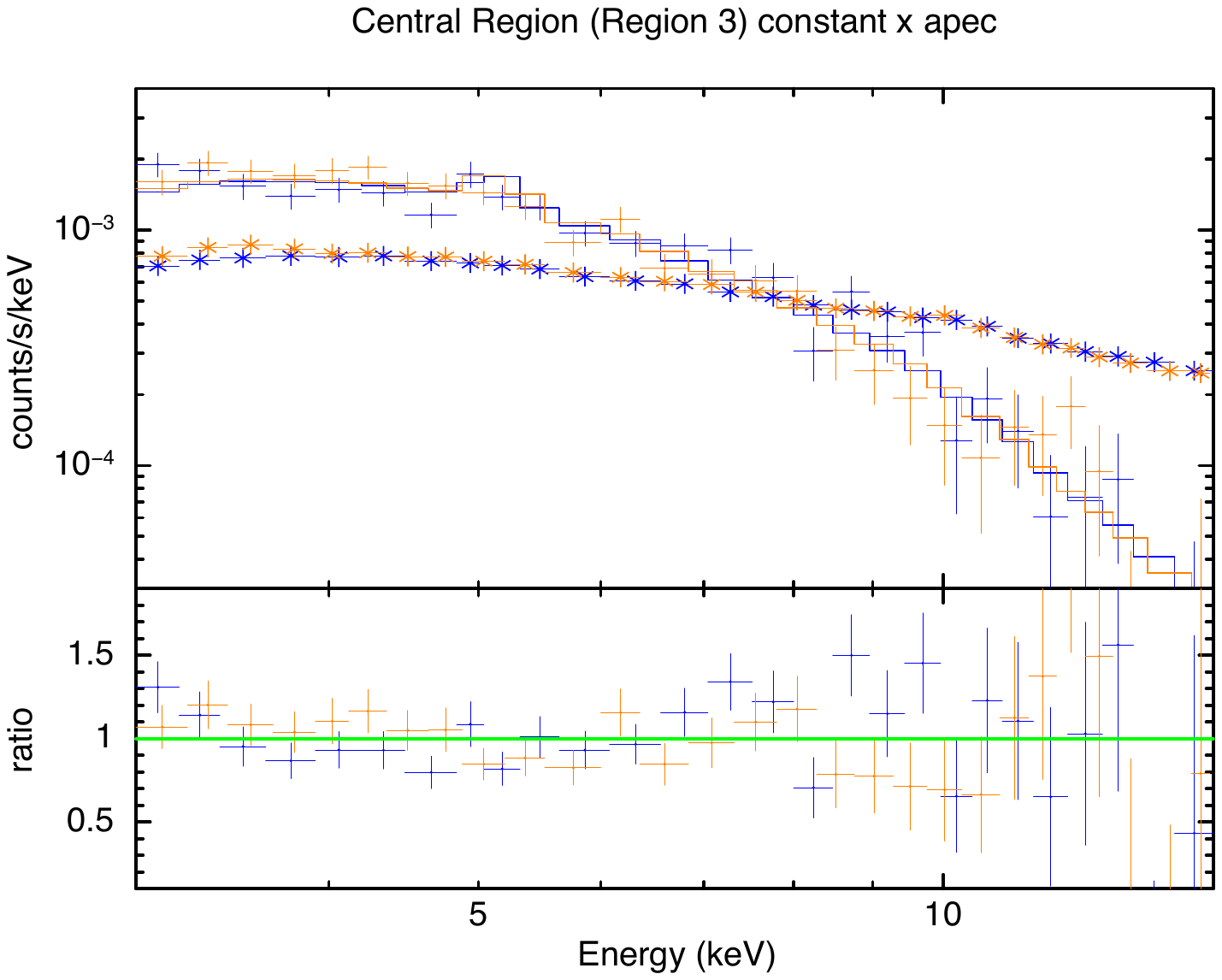}
\includegraphics[width=96mm]{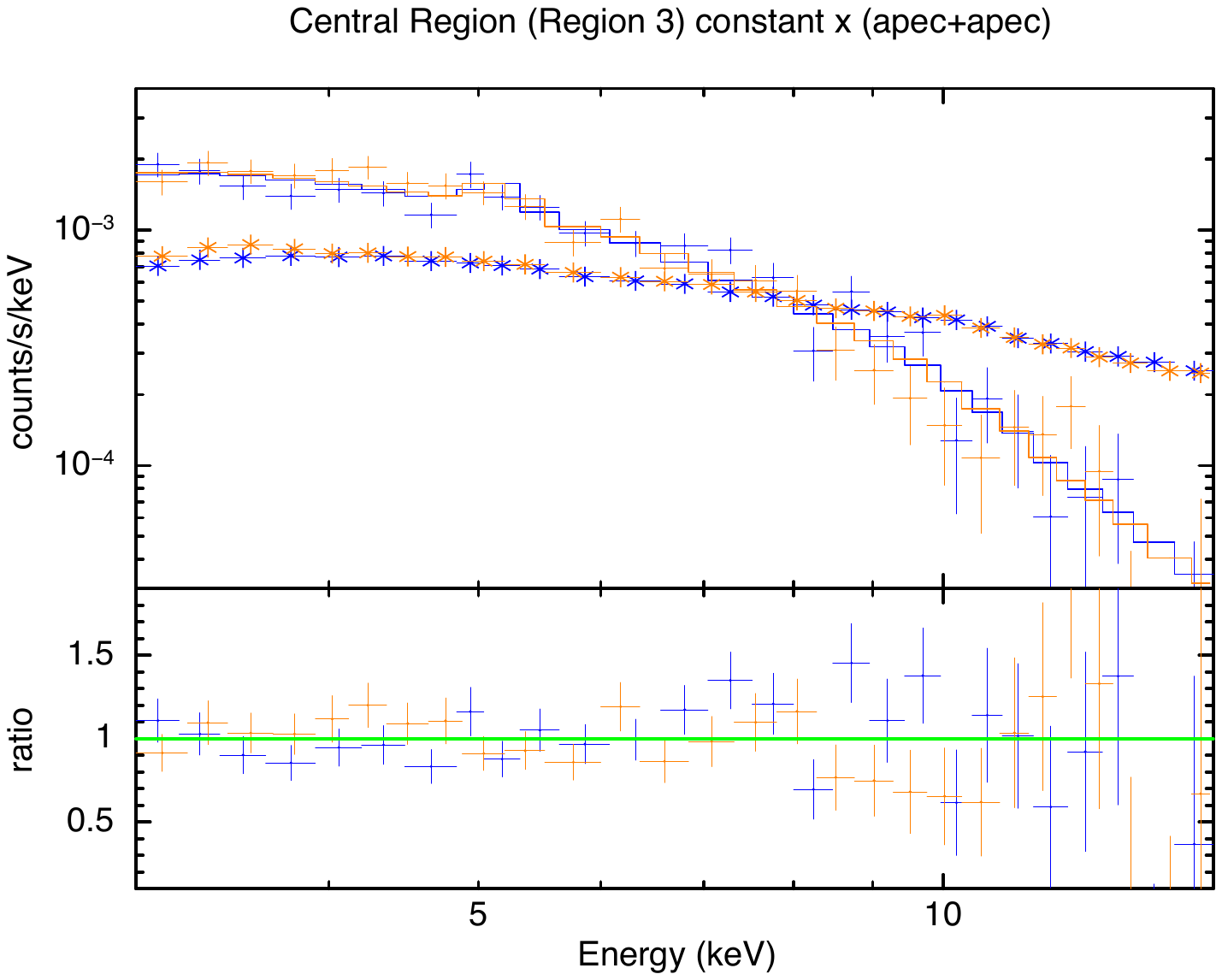}
\caption{Central region spectra showing \nustar~fits using a single and two temperature models. Background is indicated with asterix symbols. For plotting purposes, adjacent bins are grouped until they have a significant detection at least as large as 8$\sigma$, with maximum 12 bins.
\label{fig:centerspectra}}
\end{figure}

When we fit the central region (Region 3) with a single, thermal {\tt apec} model, we find a temperature of \textit{kT}=$5.46\pm0.29$~keV with a C-stat/d.o.f=493.33/537. We note that despite the statistics, residuals show less than a good fit as shown in Fig.~\ref{fig:centerspectra}.
Adding another {\tt apec} model to investigate a two temperature structure and/or any scattered emission from neighboring regions, we find the hotter component to be \textit{kT}=$6.05^{+0.41}_{-0.40}$~keV (\textit{F}$_{3-15~keV}$=2.91~$\times$~10$^{-13}$~erg/s/cm$^2$/keV.) and the cooler component to be \textit{kT}=$0.71^{+0.37}_{-0.33}$~keV (\textit{F}$_{3-15 keV}$=1.56~$\times$~10$^{-14}$~erg/s/cm$^2$/keV) with C-stat/d.o.f=485.09/535.
Given the low flux, the secondary {\tt apec} emission only constitutes about 5\% of the total emission, which provides hints that this region is slightly affected by ``cross-talk" or other systematics rather than hosting this extremely cool component, which \nustar~is expected not to be sensitive to.

For each ROI, we used \textsc{nuproducts} to extract spectra and \textsc{nuskybgd} to generate background spectra based on fits to the background across the FOV as outlined in Section~\ref{sec:reduction}. 
Spectra were then fitted using {\tt XSPEC}\footnote{\url{https://heasarc.gsfc.nasa.gov/docs/xanadu/xspec/}}. We applied modified Cash Statistics (\textbf{XSPEC} C-stat), a maximum likelihood-based statistic proposed by \citet{cash79} that is appropriate for the data and the background spectra as explained in detail by \cite{wik14}, Section 4.2. Photon counts were grouped to have at least 3 counts per bin in the spectral analysis. We used only data in the 3.0~--~15.0~keV band of \nustar's bandpass as the background dominates at higher energies.

The best-fit parameters of the models that best represent the emission from these regions are then used as a starting point value for the cross-talk fitting. All regions between 1-13 (except for Region 7) were best described by a thermal emission model {\tt apec}, whose redshift and metal abundance parameters were fixed at z~=~0.304, and \textit{Z$_{1}$}~=~0.3~{\it Z$_{\odot}$}, respectively. 

The point sources (dashed circles in Fig.~\ref{fig:ROI}) are described best by a single power-law model. We use the abundance table of Wilms \citep{wilms00}. We note that adding an absorption model with N$_{H}$=4.45$\times$10$^{20}$~cm$^{-2}$ \citep[based on Leiden/Argentine/Bonn Galactic HI survey][]{kalberla05} changes the temperature values only 1-2 percent, therefore in the cross-ARF fitting, we did not use an absorption model to simplify the process.

The Region 7 {\tt apec} model temperature was found to be around 20~keV and was unconstrained. Therefore, for this region, we used an additional {\tt powerlaw} model to account for possible AGN activity. 

This preliminary fitting procedure provides the much-more-complicated cross-talk analysis with reasonable initial values that makes the next procedure both time efficient and less likely to diverge from the global minima. We do not display the initial fit parameters as they are inferior to the \textsc{nucrossarf} fit. We refer to \citet{tumer23} for an example of how \textsc{nuproducts} and \textsc{nucrossarf} fit values compare for another merging galaxy cluster.

\subsection{Tackling the Cross-Talk: \textsc{nucrossarf}}\label{sec:cross-talk}

Due to the moderate ($\sim$1$\arcmin$ Half Power Diameter, $\sim$18$\arcsec$ Full Width at Half Maximum), slightly energy-dependent PSF of \nustar, cross-contamination, namely {\it cross-talk}, of multiple emission features in the regions of interest are produced. The standard \nustar~pipeline science products generation tools, \textsc{nuproducts}, produces ARFs for point or diffuse sources only for the user defined ROIs that does not account for the ARFs for other sources whose emission originates outside these extraction regions. In this paper we refer to those as ``cross-ARFs". 

\textsc{nucrossarf}\footnote{\url{https://github.com/danielrwik/nucrossarf}} is a set of IDL routines designed to account for this contamination by divorcing the contamination and source emission. This method is explained, implemented and tested by \citet{tumer23}, and in \citet{Creech2024} the systematic accuracy of the method is demonstrated at the few percent level. 
In addition, \citet{tumer24} use this method for \zwcl~data with ObsID 70801003002. Due to the short exposure time, hence low S/N, \citet{tumer24} applies the \textsc{nucrossarf} method to only four large ROIs, yet the approach and the method set up is identical.

\movetabledown=50mm
\begin{deluxetable*}{l|cccc|cccc}
\tabletypesize{\scriptsize}
\tablewidth{0pt} 
\tablecaption{Spectral fit results obtained from the cross-talk analysis of the \nustar~spectra from the regions show in Figure~\ref{fig:crossarffit} in the 3.0-15.0~keV band. The {\tt apec} normalization ({\it norm}) is given in $\frac{10^{-14}}{4\pi \left [D_A(1+z) \right ]^{2}}\int n_{e}n_{H}dV$. We emphasize that ``Center" indicates Region 3 and \textit{not} the X-ray centroid.
\label{tab:crossarffit}}
\tablehead{
\colhead{} & \multicolumn{4}{c}{CrossTalk Fit (\textsc{Main Fit})} & \multicolumn{4}{c}{CrossTalk Fit w/ Po in Relic} \\[-0.95em]
\colhead{C/$\nu$} & \multicolumn{4}{c}{6619.86/6604} & \multicolumn{4}{c}{6618.96/6602}
\\
\hline
\colhead{} &  \colhead{kT} & \colhead{norm} & \colhead{$\Gamma$} & \colhead{$\kappa$} & \colhead{kT} & \colhead{norm} & \colhead{$\Gamma$} & \colhead{Flux}
\\
\colhead{Region} &  \colhead{(keV)} & \colhead{(10$^{-3}$~cm$^{-5}$)}  & \colhead{} & \colhead{(10$^{-6}$)} &\colhead{(keV)} & \colhead{(10$^{-3}$~cm$^{-5}$)} & \colhead{} & \colhead{(10$^{-14}$ erg/cm$^{2}$/s})}
\startdata
\\[-0.95em]
Region 1 & 2.59$^{+0.40}_{-0.37}$& 1.409$^{+5.034}_{-3.323}$ & \nodata & \nodata & 2.60$^{+0.41}_{-0.37}$ & 1.397$^{+0.487}_{-0.321}$ & \nodata & \nodata  \\ 
\\[-0.5em]
Region 2 & 1.12$^{+0.25}_{-0.21}$& 4.871$^{+6.311}_{-2.465}$ & \nodata & \nodata& 1.13$^{+0.25}_{-0.21}$ & 4.727$^{+6.025}_{-2.024}$ & \nodata & \nodata  \\ 
\\[-0.5em]
Region 3 (Center) & 4.09$^{+1.34}_{-0.43}$ & 1.050$^{+0.185}_{-0.292}$  & \nodata & \nodata&4.11$^{+0.83}_{-0.42}$ & 1.043$^{+0.179}_{-0.202}$ &\nodata & \nodata \\
\\[-0.5em]
Region 4 & 8.30$^{+2.48}_{-1.37}$& 0.211$^{+0.035}_{-0.037}$ & \nodata & \nodata&8.32$^{+2.50}_{-1.32}$ & 0.211$^{+0.035}_{-0.037}$ & \nodata & \nodata  \\ 
\\[-0.5em]
Region 5 & 7.30$^{+13.28}_{-6.26}$& 0.056$^{+1.461}_{-0.027}$ & \nodata & \nodata&7.26$^{+13.21}_{-6.05}$ & 0.056$^{+0.129}_{-0.027}$ & \nodata & \nodata  \\ 
\\[-0.5em]
Region 6 & 7.60$^{+2.37}_{-1.59}$& 0.160$^{+0.042}_{-0.032}$ & \nodata & \nodata&7.55$^{+2.26}_{-1.57}$ & 0.160$^{+0.042}_{-0.032}$ & \nodata & \nodata  \\ 
\\[-0.5em]
Region 7 & 1.29$^{+0.72}_{-0.63}$ & 1.171$^{+10.101}_{-0.842}$ & 1.27$^{+0.66}_{-1.27}$ & 5.274$^{+17.640}_{-2.925}$ & 1.22$^{+0.72}_{-0.56}$ & 1.250$^{+12.250}_{-0.897}$ & 1.36$^{+0.59}_{-1.32}$ & 6.05$^{+15.9}_{-3.27}$  \\ 
\\[-0.5em]
Region 8 (NR$_{in}$) & 9.90$^{+2.81}_{-2.55}$& 0.147$^{+0.037}_{-0.023}$  & \nodata & \nodata&9.50$^{+2.91}_{-2.15}$ & 0.152$^{+0.033}_{-0.023}$ & \nodata & \nodata \\
\\[-0.5em]
Region 9 (NR) & 3.44$^{+16.46}_{-2.83}$ & 0.065$_{-0.056}^{+0.699}$   & \nodata & \nodata&1.00$^{+4.72}_{-0.58}$ & 1.050$^{+42.832}_{-1.242}$ & 1.87 & 0.70$^{+1.09}_{-0.70}$ \\
\\[-0.5em]
Region 10 (NR$_{out}$) & 1.76$^{+1.15}_{-0.47}$ &0.922$^{+1.402}_{-0.617}$ & \nodata & \nodata & 1.97$^{+0.84}_{-0.57}$ & 0.711$^{+1.042}_{-0.375}$ & \nodata & \nodata  \\ 
\\[-0.5em]
Region 10 \tablenotemark{$\ast$} (NR$_{out}$) & 1.01$^{+0.05}_{-1.01}$ &2.986$^{+0.663}_{-0.478}$ & \nodata & \nodata & \nodata & \nodata & \nodata & \nodata  \\ 
\\[-0.5em]
Region 11 (SR$_{in}$) & 5.70$^{+1.13}_{-0.97}$ & 0.387$^{+0.089}_{-0.067}$  & \nodata & \nodata &  5.29$^{+1.32}_{-0.75}$ & 0.414$^{+0.084}_{-0.086}$ &\nodata & \nodata \\ 
\\[-0.5em]
Region 12 (SR) & 1.42 $^{+0.52}_{-0.27}$&2.109$^{+2.264}_{-1.211}$  & \nodata & \nodata & 1.24$^{+0.48}_{-0.27}$ & 2.811$^{+3.841}_{-1.690}$ & 1.97 & 1.41$^{+1.64}_{-1.41}$\\
\\[-0.5em]
Region 13 (SR$_{out}$) & 2.21$^{+0.61}_{-0.46}$&1.176$^{+0.861}_{-0.451}$  & \nodata & \nodata & 2.17$^{+0.58}_{-0.43}$ & 1.225$^{+0.713}_{-0.464}$ &\nodata & \nodata \\ 
\\[-0.5em]
Region 13 \tablenotemark{$\ast$}(SR$_{out}$) & 2.22$^{+0.19}_{-0.27}$&1.131$^{+1.133}_{-1.131}$  & \nodata & \nodata & \nodata & \nodata &\nodata & \nodata \\
\\[-0.5em]
\enddata
\tablenotetext{\ast}{These values are obtained from repeating the main spectral fit where the energy band is restricted to the 3.0-8.0~keV band for region Region 10 and Region 13 alone and Region 10 (13) background is varied by -5\%/15\% (15\%/-20\%) for FMPA/B, respectively. The resulting statistics is C-stat/d.o.f.=6038.30/5996.}
\end{deluxetable*}

\begin{figure*}
\centering
\includegraphics[scale=0.58]{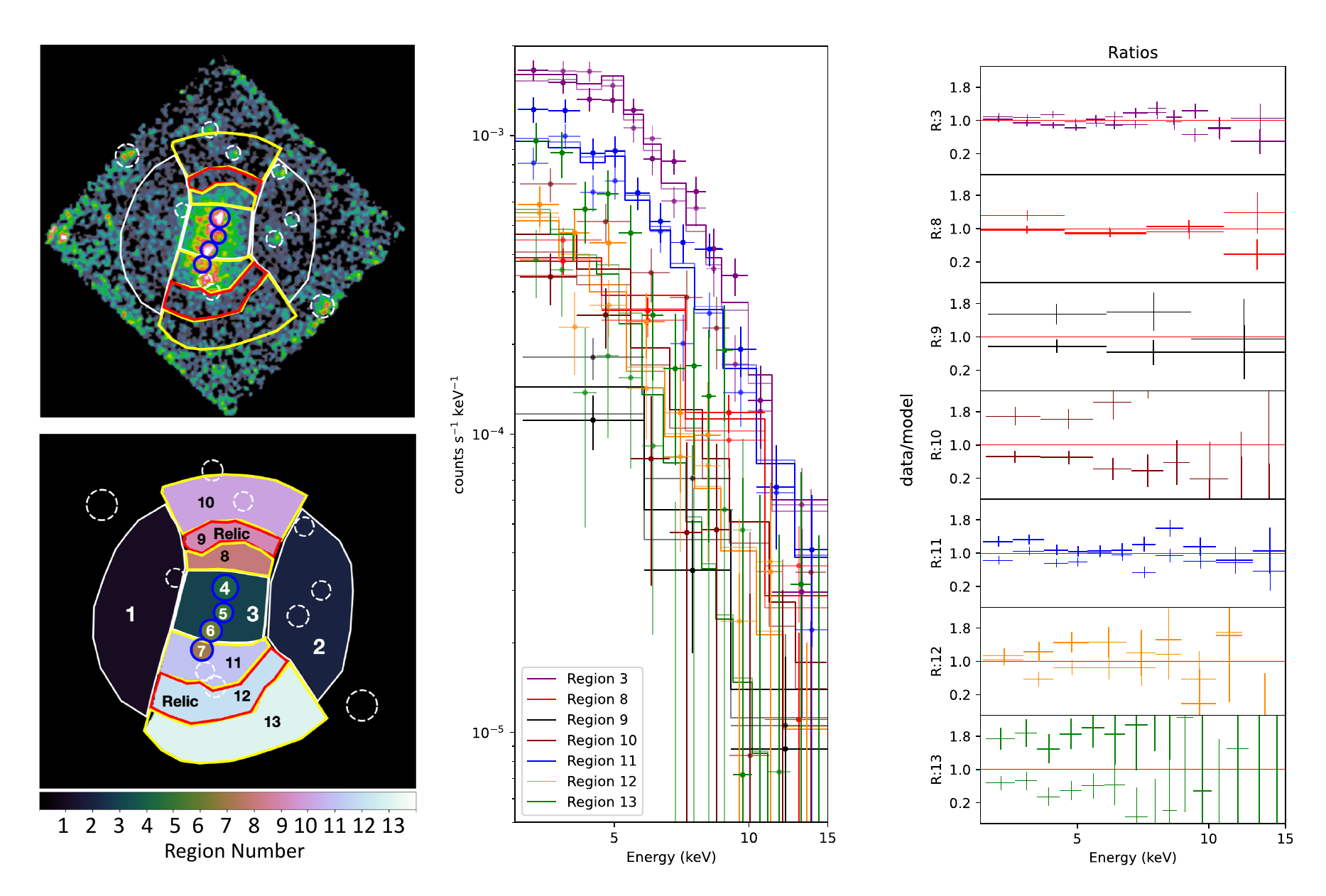}
\caption{Regions of interest figure (Fig~\ref{fig:ROI} is duplicated here to facilitate visualization of region vs spectrum correspondance.  Joint cross-talk spectral fits from all regions ({\it Middle Panel}). 
A model combining both FPMA and FPMB observations (solid) is compared to combined data from both FPMs (points) for each region. Each observation was fit individually, this is only for illustrative purposes. Each model and data comparison is vertically shifted logarithmically to better distinguish between regions ($\times 10^{shift}$). 
For each region the quality of the fit is represented as the ratio:data/model ({\it Right Panel}). The y-axis lists region numbers, corresponding with region numbers on the left. 
\label{fig:crossarffit}}
\end{figure*}

We provide a visual of the PSF cross-talk made available by the \textsc{nucrossarf} code in the Appendix section of this paper. Such visualization helps highlight the cross-contamination. 

Using the model parameter values from the preliminary analysis---directly fitting the NuSTAR spectra without accounting for the cross-talk---we set up two sets of \textsc{nucrossarf} runs. The difference between the first and second runs is the additional {\tt powerlaw} model to represent any non-thermal (IC) emission coming from the relic regions. We refer to the first run as ``\textsc{Main Fit}".

\begin{figure}
\centering
\includegraphics[width=88mm]{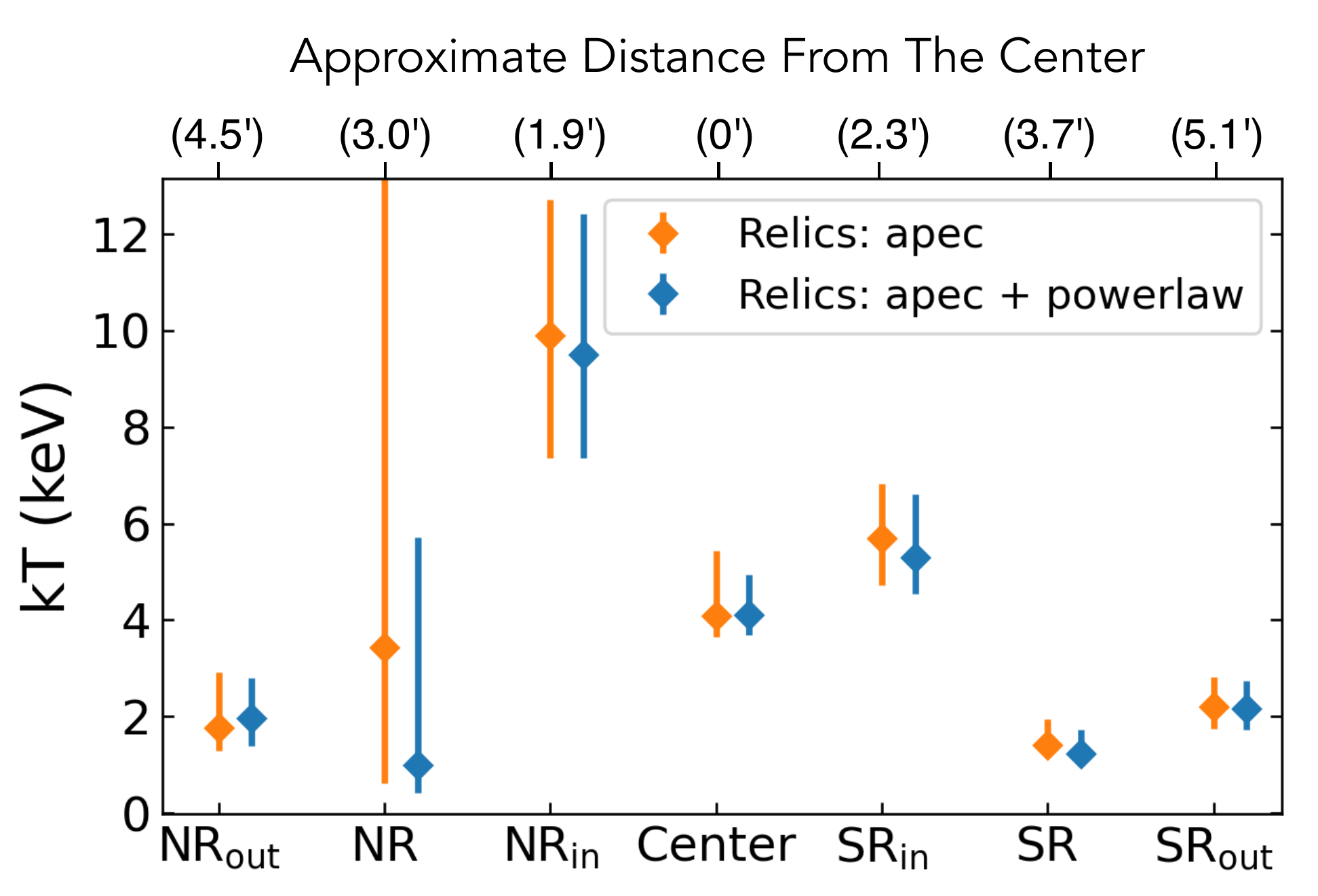}
\caption{Best-fit temperature profile for the cluster plotting the results shown in Table~\ref{tab:crossarffit}. Orange data represents \textit{kT} values and corresponding errors from the \textsc{Main Fit} shown in Table~\ref{tab:crossarffit} and the blue represents the fit results where the relics were modeled with an additional {\tt powerlaw} component. Upper x-axis provides approximate angular distance of each region center with respect to Center region in arcminutes yet the tick marks are equidistant. We emphasize that ``Center" indicates Region 3 and \textit{not} the X-ray centroid.
\label{fig:kTprofile}}
\end{figure}

In addition to the 13 diffuse regions, we modeled 10 point sources indicated by the Chandra analysis \citep{tumer24} with a {\tt powerlaw} model. We fixed the photon indices of these sources to the values obtained from \textsc{nuproducts}, but their normalizations were freed to vary. We do not expect the indices to change and leaving them free not only results in much longer computational time but also model parameters of these weaker sources can easily take on unphysical values, unrealistically mix with components from other regions, and in any case be poorly constrained.

As in the \textsc{nuproducts} run, the {\tt apec} model redshift and metal abundance values were fixed to $z=0.304$ and $Z=0.3~Z_\odot$, respectively. During this procedure, the {\tt apec} temperature, abundance and redshift parameters were linked between the modules (FPMA and FPMB). We accounted for the cross-calibration differences of detectors with the {\tt constant} model, where the normalization of the models of different instruments were tied. The joint-fit model hence becomes; {\tt constant}$\times$~{\tt apec}. We fixed the constant to 1, and the constant was fixed to 0.97 for all regions and cross-ARF models belonging to FPMB, which was obtained from the Central region spectra fit. Given the number of regions and and cross-ARFs, one could only keep one of the constant parameters (a nuisance parameter) and leave the rest free, constant parameter alone would increase the d.o.f. by 23x23-1=528.

To investigate the contribution from potential IC emission, we fixed the photon index to radio index values $\Gamma$~=~1.87 for NR and $\Gamma$~=~1.97 for SR, derived from $\alpha_{inj}$ reported by \citet{jones21} ($\Gamma = \alpha+1$, for the standard definitions $F_X\propto E^{-\Gamma}$ and $F_R \propto \nu^{-\alpha}$) for the second run of \textsc{nucrossarf}. Here we assume that the electron population producing the radio synchrotron emission is also responsible for upscattering CMB photons to the X-ray regime. When we let these indices free to vary, we find that neither indices were constrained. In both relic regions, the presence of IC emission is only significant at the $\sim$1$\sigma$ level, insufficient for a claim of detection.

We also refit the cross-ARF spectra and let the abundance and redshift parameters of the {\tt apec} model free. We note that the temperature changes by 3 percent for a 4 keV plasma, and for regions with low S/N (Region 10, 13), the abundances hit the upper value of solar, but for other regions, lie in between 0.3-0.4 Z$_{solar}$ with $\Delta$C-stat/$\Delta$d.o.f.= 30.94/26. 

The fit results are shown in Table~\ref{tab:crossarffit}. In Figure~\ref{fig:crossarffit}, we show spectra and models from the \textsc{Main Fit}. This plot displays only the 7 regions along the merger axis for clarity, where the rest of the 6 regions as well as 10 point sources are not shown as plotting all spectra hides the spectral features of the main ROIs. An example of how the complete set of spectra from a large number of regions look like is displayed in \citet{tumer23}.

We also plot the results for the Center, NR, SR and corresponding post/pre-shock regions to visualize a cluster temperature profile from North to South in Fig.~\ref{fig:kTprofile}. 

Since the FMPA and FPMB data/model ratios diverge for Regions 10 and 13 as seen in Figure~\ref{fig:crossarffit} (right panel), we refitted these these spectra by varying the \textsc{nuskybgd} background using the \textsc{corfile} task while keeping the rest of the region parameters fixed to those found in the \textsc{Main Fit} where the relics are modeled with single {\tt apec} models only, indicated as ``\textsc{Main Fit}" in Table~\ref{tab:crossarffit}. 

\begin{figure}
\centering
\includegraphics[width=85mm]{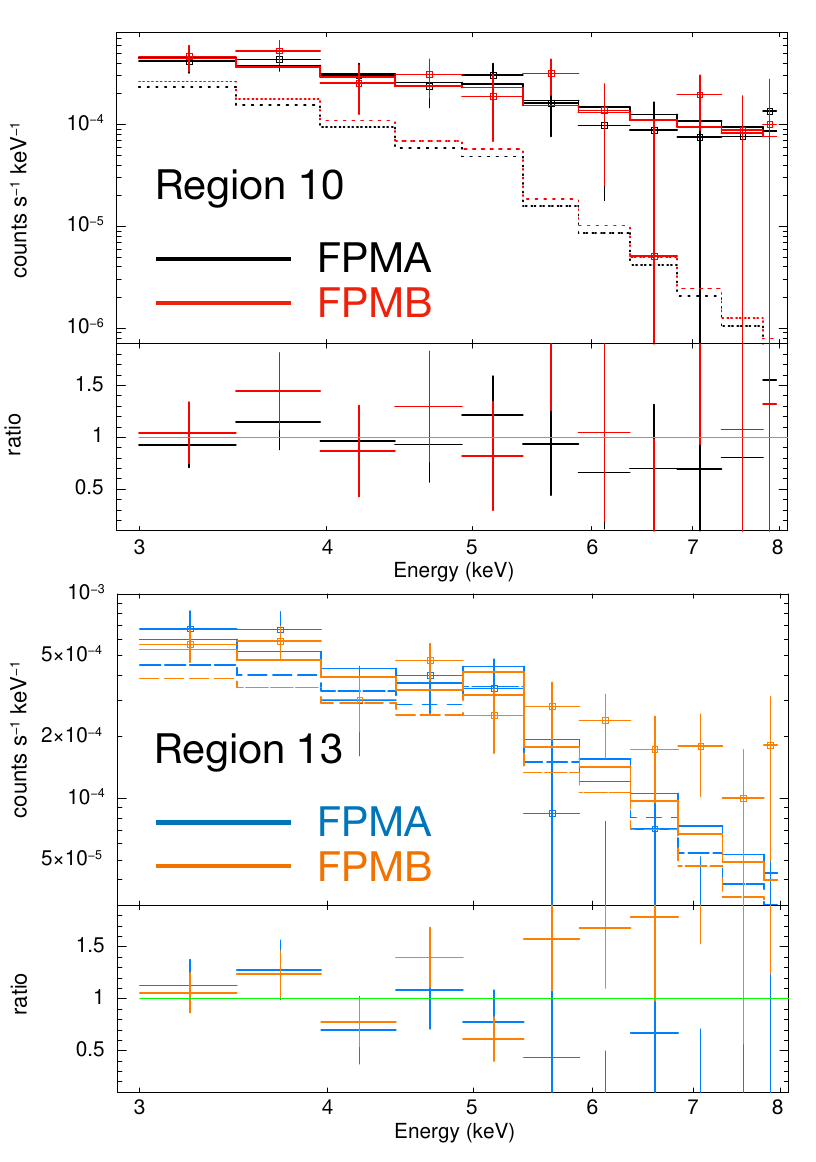}
\caption{Spectral fit for Regions 10 and 13 after background variation and energy band restriction are applied. Solid lines represent the total model, all ARF contributions, and dotted/dashed lines the model convolved with only the intrinsic ARF; the ARF of the region of interest. The ratio panels represent the ratio of data to the model.
\label{fig:Reg10n13fit}}
\end{figure}

Since these regions point to cooler temperatures, we ignored the data beyond 8.0~keV to further eliminate the effect of background variations. This is aimed at better investigating the parameters for these regions since the temperature of these regions with S/N dictate the accuracy for shock strength. While the temperature and normalization fit results are given in Table~\ref{tab:crossarffit}, Region 10 (13) background is varied by -5\%/15\% (15\%/-20\%) for FMPA/B, respectively. These values are obtained by visually inspecting the ratio with various percentages, and selecting the one that provides a ratio closest to one in the 3.0-6.0~keV energy band. Beyond 6.0~keV, the background variation does not remedy the ratios since the count rate drops exponentially whereas the background is mainly flat. The resulting spectra are plotted in Figure~\ref{fig:Reg10n13fit} by individual selection of Region 10 and 13 components alone and the fit values are reported in Table~\ref{tab:crossarffit} at rows indicated by an asterix, with the statistics noted in the footnote. The statistics of the joint fit became C-stat/d.o.f.=6038.30/5996 which gives $\Delta$C-stat/$\Delta$d.o.f.=581.56/608 with respect to the \textsc{Main Fit}, where the change in d.o.f. is due to eliminating data beyond 8~keV. 
The detailed spectra for each module of the each region is shown in the Appendix section to present the dominant background emission in these regions.

Moreover, we tied the temperature of Region 9 to Region 8, as well as temperature of Region 12 to Region 11, for a configuration where the post-shock regions are  Region 8 + Region 9 for North, and Region 11 + Region 12 for South. 

This fit yielded a Northern post-shock temperature of \textit{kT}=11.26$^{+1.54}_{-1.40}$~keV
and pre-shock (Region 10) temperature of \textit{kT}=1.04$^{+0.05}_{-1.04}$~keV. In the South post-shock temperature was \textit{kT}=4.74$^{+0.21}_{-0.23}$~keV and pre-shock (Region 13) temperature was \textit{kT}=1.56$\pm{0.11}$~keV
This new fit improved the statistics by $\Delta$C-stat/$\Delta$d.o.f.=29.41/6. Although the Southern Mach becomes similar ({$\mathcal{M}$}$\sim$2.6) to what was found for the \textsc{Main Fit}, the Northern shock strength seems to be larger ({$\mathcal{M}$}$\sim$5.5) for this configuration. We note, as shown above, the unbound lower limit of \textit{kT} for Region 10.

The statistics from the \textsc{nucrossarf} fits with and without additional {\tt powerlaw} models applied to relic spectra are very similar. Due to the marginal presence of IC components, we use temperature results from the \textsc{main fit} fit without this addition to calculate the Mach number and the errors. Uncertainties are propagated assuming symmetric Gaussian distributions in the standard way, which gives the following expressions with upper and lower limits calculated separately:

\begin{equation}
\Delta (T_2/T_1)= \frac{T_2}{T_1} \sqrt{(\Delta T_2/T_2)^2 + \Delta T_1/T_1)^2}
\label{eq:mach_error}
\end{equation}

\begin{equation}
\Delta \mathcal{M} = \frac{4\mathcal{M}}{\sqrt{8(T_2/T_1-7)^2+15}}\Delta (T_2/T_1)\\
\label{eq:mach_error2}
\end{equation}
\\

where $T_2$ and $T_1$ denote the post- and pre-shock temperatures, respectively, and low.

Furthermore, we explored the parameter covariances between pre/post-shock temperatures to test their implications on the Mach number. In the joint crossarf fit, we fixed the NR$_{in}$, pre-shock temperature to the upper limit (12.5~keV) and ran the full fit including the powerlaw model in the relics and the resulting NR$_{out}$ resulted in 2.00~keV temperature, a $\leq$2\% increase from the baseline. The other region temperatures changed at most 2\%, except for Region 5 that increased 15\%, whose uncertainty is the largest among all regions due to poor statistics and multiple adjacent regions that contribute to cross contamination, and the lowest flux region with 2.8$\times$10$^{-14}$ erg/cm$^{2}$/s in the 3.0-15.0~keV band. The 15\% increase is within the 1$\sigma$ errors of what is reported in Table~\ref{tab:crossarffit}. We chose the Northern shock since it is where we report the unusually large Mach number.

\section{Discussion}\label{sec:discussion}

\subsection{Cluster Center}
Prior to the crosstalk fitting (Section \ref{sec:cross-talk}), we fit the central region (Region 3) excluding the BCGs (regions 4 and 6) as well as Region 5, which is likely an X-ray tail stripped from the Northern BCG similar to what is seen in the merger of MACS1752 \citep{finner21}. 

We considered two different models to describe the central emission; a single {\tt apec} model and a two {\tt apec} model during the assessment of initial values to be used by \textsc{nucrossarf}. The single {\tt apec} model indicated a plasma with a temperature of kT=5.46 keV, yet the data/model ratio diverged from unity with features at different energies. We then added a second {\tt apec} model to accommodate a multi-temperature plasma and/or any PSF scattering from regions 4, 5, and 6. While the hot component remained within 1$\sigma$ of what was found with the single {\tt apec} model, the cooler temperature was 0.71 keV with a  $\Delta$C-stat/$\Delta$d.o.f=8.24/2. 

At such low temperatures, the emission would be too faint to be detected in NuSTAR's hard X-ray band. In order for it to be detectable, the emission would have to be very bright at softer energies, in which case it would have already been observed by soft X-ray observatories. Therefore, including this component is not physically justified.
In addition, the cooler component accounts for less that 5\% of the total emission flux, we conclude that the improved statistics point to a scenario where the secondary {\tt apec} model is simply accounting for contamination from neighboring regions and/or background fluctuations.

Spectral fitting of the central region while excluding the BCGs with the \textsc{nucrossarf} method resulted in a well constrained $\simeq$4 keV plasma, where regions 4, 5, and 6 show much higher temperatures (Table~\ref{tab:crossarffit} and Fig.~\ref{fig:kTprofile}). The cluster merger stage seems to be at the returning phase, $\sim$1.7 Gyr after first pericenter passage \citep{finner21}. 
Similar late-stage merger scenarios have been inferred in other systems. The low-mass cluster PSZ2 G181.06+48.47 has exceptionally distant radio relics, suggesting shocks that propagated far into the outskirts, consistent with a merger well past first pericenter passage \citep{stroe25}.
Given the hot temperature of Region 5 (comparable to both subclusters found in this paper), enhanced surface brightness in \xmm~images, a weak lensing peak lying to the south of the Northern clump, and an indication of a radio source \citep{finner21}, it is possible that this merger is more complicated than the originally assumed scenario of a 1:1 mass ratio head on collision of two clusters. It is possible that the Southern subcluster had already been going through a minor merger before encountering the Northern subcluster.

\subsection{Post-shock and pre-shock regions temperatures}

We find significant temperature jumps across the relic regions, which were selected to match the extent of the radio emission. The temperature jump is much stronger across the NR than the SR. While the pre-shock regions at both shock fronts show similar temperatures, the post-shock region, Region 8, in the NR appears to be extremely hot. The post-shock region definition where the Region 8 and Region 9 temperatures are tied reveal an even hotter post-shock plasma (\textit{kT}=11.26$^{+1.54}_{-1.40}$~keV).

This supports the merger scenario of a slightly tilted merger axis: i.e., the SR-center axis is tilted away from the observer, as supported by the Southern relic's rotated polarization angle \citet{deGasperin14}. We view the Northern shock edge-on, hence the high X-ray Mach number we find, also hinted by the perpendicular electric field vectors with respect to the relic extension, assuming that the magnetic field is  aligned with the relic extension \citep[see, e.g.,][]{weeren10}.
The larger spatial extent of the Southern relic may also point to a multiple merger, evidenced by the low frequency radio observations \citep[LOFAR;][]{jones21} that reveal a larger spatial extent of the relic with respect to its high frequency counterpart \citep[WSRT;][]{deGasperin14}.
The coincidence of radio relics with X-ray shocks has been highlighted in other systems, such as the Bullet Cluster, where \cite{shimwell15} identified a periphery shock coincident with a radio relic and argued that it is powered by shock re-acceleration of seed electrons from a nearby radio galaxy remnant. Taken together, these results strengthen the interpretation that the NR and SR in our system trace merger-driven shocks that both heat the ICM and re-accelerate pre-exisiting relativistic electrons.

The pre-shock region spectral fitting for both relics, namely regions 10 and 13, show variations for FPMA and FPMB as seen in Fig.~\ref{fig:crossarffit}. We attribute these to imperfect background modeling that becomes significant at these low S/N values. 

The temperature in both of these regions are well constrained around $\sim$2~keV, yet the data/model ratios diverge in the \textsc{Main Fit} shown in Fig.~\ref{fig:crossarffit}. Therefore, we further investigated this discrepancy. We varied the background and constrained the spectra to 3.0-8.0~keV to be able to constrain the temperatures better. Although the SR$_{out}$ region temperature did not change from the \textsc{Main Fit} results, the NR$_{out}$ region temperature dropped even further to plasma temperatures that \nustar~is not particularly sensitive to. However, with a much higher S/N, the large Region 2 to the West has a similar ambient temperature and is well constrained. With the data at hand, it is difficult to assert with high confidence that a physical emission from the NR$_{out}$ is detected. Regardless, assuming a temperature of this region around 2~keV is a conservative approach given that Region 1 and Region 2 temperatures lie within 1.1-2.6~keV. While 1--2~keV temperatures may seem beyond \nustar's reach given its $E>3$~keV sensitivity, in fact the temperature is {\it easier} for \nustar\ to constrain since the emission comes from the steeply declining exponential tail of the bremsstrahlung spectrum. In very long observations such as this one, low temperature emission can be detected and constrained even when comparable to the background level.

\begin{deluxetable}{lccc}
\tabletypesize{\scriptsize}
\tablewidth{0pt} 
\tablecaption{Mach numbers of the candidate X-ray shock fronts calculated from the temperature jumps obtained from the \textsc{main fit}.
\label{tab:mach}}
\tablehead{\\[-0.95em]
&\multicolumn{2}{c}{Region Temperature (keV)}& {$\mathcal{M}$}\\[-0.5em]
&\colhead{{post-shock}}& \colhead{{pre-shock}} & Eqn.s~\ref{eq:mach_temp},\ref{eq:mach_error},\ref{eq:mach_error2}}
\startdata
\\[-0.95em]
{\nustar}& N$_{in}$ (9.90$^{+2.81}_{-2.55}$) & N$_{out}$ (1.76$^{+1.15}_{-0.47}$) & $3.90^{+1.64}_{-0.85}$ \\  
&S$_{in}$ (5.70$^{+1.13}_{-0.97}$) & S$_{out}$ (2.21$^{+0.61}_{-0.46}$)& $2.36^{+0.58}_{-0.46}$ \\
\\[-0.95em]
\hline
\\[-0.95em]
\\[-0.95em]
\enddata
\end{deluxetable}

\subsection{Shock Strength}

Using the temperatures of the post-shock and pre-shock regions found from the \textsc{nucrossarf} spectral analysis (Section \ref{sec:cross-talk}) and applying Rankine-Hugoniot conditions (Eqn.~1), we calculated the Mach numbers associated with the radio shocks (relics).
The Mach numbers are shown in Table \ref{tab:mach}, 
which use temperatures obtained from the \textsc{\textsc{main fit}} analysis that does not include an additional {\tt powerlaw} model for the relic regions (since the addition of a {\tt powerlaw} component is not justified statistically.
We report $\mathcal{M}$=$3.90^{+1.64}_{-0.85}$ and $\mathcal{M}$=$2.36^{+0.58}_{-0.46}$ for the NR and SR, respectively.
Radio derived Mach numbers for the Northern and Southern shocks are $\mathcal{M}$=2.5$\pm$0.2 and $\mathcal{M}$=2.3$\pm$0.2, respectively \citep{jones21}. 

Although for the SR, the X-ray and Radio Mach numbers agree within 1$\sigma$, for the north relic, the Mach numbers differ by a little bit more (1.6~$\sigma$). Interestingly, the X-ray shock strength is at least 50\% stronger than the Radio shock strength. This finding is rare given that radio observations are more likely to capture regions with high cosmic-ray acceleration \citep{lee25,botteon20}. Given that the Mach number is averaged over the shock surface, it is possible that much stronger shocks exist within the shocked area. That said, the discrepancy has a low significance and should not be over-interpreted.

Since the shock strength depends on constraining the location and the temperature of the regions behind and ahead of the shock, we applied multiple approaches to enhance our claim. 
Given that the \textsc{Main Fit} residuals were suspicious for the pre-shock regions, we repeated the \textsc{Main Fit}; the full \textsc{nucrossarf} joint fitting procedure, by narrowing the spectral band range of Region 10 and Region 13 and varied the background by $\sim$5\% increments up to 20\%, allowed by the few percent systematic uncertainty in the background reconstruction and the fact that the signal lies below the background level (see Appendix~\ref{sec:app:c}). This fit with $\Delta$C-stat/$\Delta$d.o.f.=581.56/608 resulted in a similar temperature for Region 13 and a lower temperature for Region 10, with respect to the \textsc{Main Fit}.

Moreover, in addition to the above alterations, we also tied the temperatures of NR and NR$_{in}$ together and tied the temperatures of SR and SR$_{in}$ to each other as well, adopting a larger region size for the post-shock regions. This fit improved the statistics by $\Delta$C-stat/$\Delta$d.o.f.$=29.41/6$, but the NR$_{out}$ temperature lower uncertainty was unbound, providing only an upper limit on the temperature there. Consequently, the Southern Mach number ({$\mathcal{M}$}$\sim$2.6) remained unchanged from the \textsc{Main Fit}, while the strength of the Northern shock increased ({$\mathcal{M}$}$\sim$5.5), albeit without good precision. 

Since the shock strength is mainly driven by the temperature of post-shock regions, we claim that we indeed observe a shock with at least $\mathcal{M}$=3 (based on the 1$\sigma$ lower uncertainty from the \textsc{Main Fit}) in the North. X-ray data of higher soft X-ray band sensitivity than that provided by \nustar\ is necessary to confirm one of the largest cluster merger shocks so far claimed in the literature. This issue is also present in the galaxy cluster CIZA J2242.8+5301, where the \xmm~systematics prevent putting strong constraints on the pre-shock temperature
\citep{ogrean13}.

We also note that the shock strengths are expected to be influenced by magnetic field turbulence around the radio relics. \cite{de_rubeis_magnetic_2024} performed a depolarization study and found that small-scale fluctuations in the magnetic field can significantly affect the observed polarization fraction and the structure of the relics. Incorporating turbulence constraints into our X-ray measurements could provide a more complete physical description of the merger shocks.

\subsection{Relic Regions}

Our joint spectral fits for the regions of interest that cover the NuSTAR field of view do not favor additional {\tt powerlaw} models at the relic sites. Adding {\tt powerlaw} models to the relic region emission, where the {\tt powerlaw} index was fixed to the radio spectral index inferred values \citep[assuming the IC and synchrotron emission are due to the same relativistic electron population]{jones21}, results in a $\Delta$C-stat/$\Delta$d.o.f.=0.90/2, as shown in Table~\ref{tab:crossarffit}. 
Assuming the photon indices can be linked to the radio non-thermal emission, non-thermal X-ray emission is detected with a significance less than 1$\sigma$, a non-detection. We therefore rule out any significant inverse Compton component in the data.  Moreover, when we leave the {\tt powerlaw} indices free during spectral fitting, the indices become unconstrained. Regardless, using the flux upper limits from Table~\ref{tab:crossarffit}, we provide a lower limit on the 20.0-80.0~keV magnetic field strength at the relics of at least 0.5~$\mu$G and 0.9~$\mu$G, for the Northern and Southern relics, respectively. This corresponds to a cluster average  magnetic field strength of $\sim$0.15~$\mu$G.

Recent work by \cite{lin_semisupervised_2024} explored the use of semisupervised machine learning to detect IC in galaxy clusters. While such an approach offers a promising complimentary method to direct spectral fitting, their performance in practical applications did not yet appear to surpass manual fitting to detect faint IC signals. Nevertheless, their approach could provide valuable comparisons for future, deeper observations.

The relics enclose a large region in the sky, therefore any IC emission component is most likely dominated by the thermal emission as well as background. It is interesting that while the statistics do not improve, the addition of a {\tt powerlaw} model to the relic regions results in better temperature constraints for all regions, most significantly for the NR region. 
This additional faint {\tt powerlaw} component may be fitting any residual unresolved background emission in the background model that becomes significant in low S/N regions such as relic locations.

While the NR temperature decreases with the additional {\tt powerlaw} component, falling below the NR$_{in}$ temperature, it is still within 1$\sigma$ of NR$_{in}$ temperature but still has large errors. We note that this region, Region 9, is the lowest flux region.
The SR region shows lower temperatures than in SR$_{in}$ with or without the additional {\tt powerlaw} component. This is not expected given that the cluster temperature profile is expected to show a gradual radial decrease. 
Therefore, we are not able to conclude whether the Northern relic site truly hosts a surface brightness depression or instead has a low temperature plasma along with IC emission, nor can we confirm whether the Southern relic site contains a low temperature plasma.
Resolving these questions requires high-angular-resolution soft-X-ray observations. 
Furthermore, the choice of relic regions in this work is motivated by the investigation of IC emission and the X-ray counterparts to the radio emission. However, the X-ray shock front is expected to lie close to the \textit{outer} edge of the radio relic emission, meaning that the NR and SR regions are likely to enclose both shocked and possibly unshocked plasma.
Once the locations of X-ray surface brightness drops are established, the relic regions could be redefined and combined with their respective inner regions to isolate the post-shock emission more effectively. 
Deep XMM-Newton observations would be able to locate the expected surface brightness drops and/or depression, and 
when used in conjunction with \nustar\ data, would have the power to detect any hard emission feature. 
In contrast, \chandra's reduced soft band response due to contamination makes it unlikely to achieve this goal.

\section{Conclusion and Future Work}
\label{sec:future}
In this work, we studied the merging cluster \zwcl~temperature structure by using deep ($\sim$300~ks) NuSTAR observations. We focused on regions corresponding to two previously detected radio relics, which are typically signatures of shocks, in this double relic cluster. 
We used \textsc{nucrossarf} to account for PSF contamination across all 13 regions of interest and 10 point sources. We find a large X-ray temperature jump across the Northern relic that corresponds to a Mach number of 3.9, that exceeds radio-determined Mach number. We find temperature jumps at the positions of both relics, indicative of shock fronts that likely generate the northern and southern radio relics, and weak, statistically inconclusive evidence for IC emission at the relic sites. Moderate \xmm\ exposure is need to constrain/distinguish the emission components at location of the radio relics.

\section{acknowledgments}

We thank the anonymous referee for their insight and suggestions. The material is based upon work supported by NASA under award number 80GSFC24M0006. DRW acknowledges support from NASA ADAP award 80NSSC19K1443. AT, CTN and DRW acknowledge support from NASA NuSTAR Guest Observer grant NNH22ZDA001N-NUSTAR.
This research has made use of data from the \nustar\ mission, a project led by the California Institute of Technology, managed by the Jet Propulsion Laboratory (JPL), and funded by by the National Aeronautics and Space Administration (NASA). In this work, we used the NuSTAR Data Analysis Software (\textsc{nustardas}) jointly developed by the ASI Science Data Center (ASDC, Italy) and the California Institute of Technology (USA). The data for this research have been obtained from the High Energy Astrophysics Science Archive Research Center (HEASARC), provided by NASA’s Goddard Space Flight Center. 


\bibliography{zwcl1856}{}
\bibliographystyle{aasjournal}

\appendix

\section{Cross-Talk}
\label{sec:crosstalk}
\begin{figure*}[h!]
\centering
\includegraphics[width=180mm]{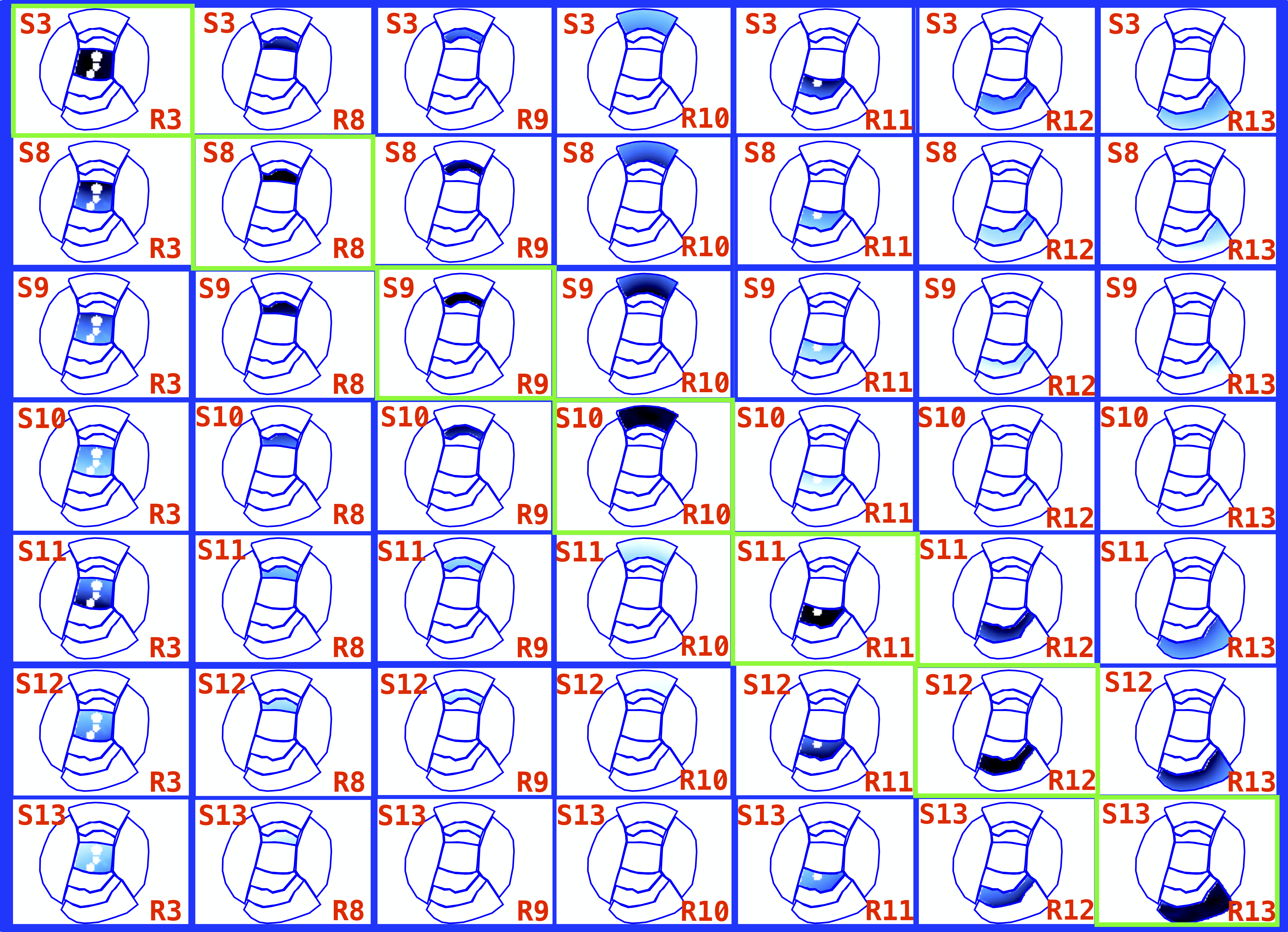}
\caption{PSFs images from the \textsc{nucrossarf} \textsc{Main Fit}, that displays the main 7 regions of interest. The intensity of the PSF contribution/contamination is in proportion with the color density, i.e.; 
sarker the shade in the figure, higher the PSF contribution/contamination. We provide a detail description of the construction of this figure in section \ref{sec:crosstalk}
\label{fig:PSFimages}}
\end{figure*}

The visual 7~$\times$~7 matrix given in Figure~\ref{fig:PSFimages} presents the mapping of the contribution from the local photons within a region and the cross-talk, where the selected regions are overlaid. We only represent the PSF images of the 7 regions of interest from whose spectral fit, the temperature profile of this cluster is plotted in Fig.~\ref{fig:kTprofile}. 
Letter--Number indicators in the boxes refer to the origin and destination of the PSF contribution. While letter ``S" point to the origin of the source from a particular region, letter ``R" indicates the region where photons land on the detector. Darker shades correspond to the amount of the PSF contained on a particular region. For instance; S10-R12 box shows how much of the fraction of the PSF Region 10 source contaminates Region 12. Therefore, the diagonal boxes highlighted by bright green show the \textit{intrinsic} emission, i.e.; the source emission filling the sky region it originated from, hence the highest fraction of the PSF (darker). This figure provides guidance on interpreting the spectral-fit results.

\section{Background Spectra}
We modeled the \nustar~background using a set of IDL routines called \textsc{nuskybgd} that has become a standard for background assessment of extended sources. The treatment of the background and its components are explained in detail in \citet{wik14}.

\begin{figure*}[h!]
\centering
\includegraphics[width=180mm]{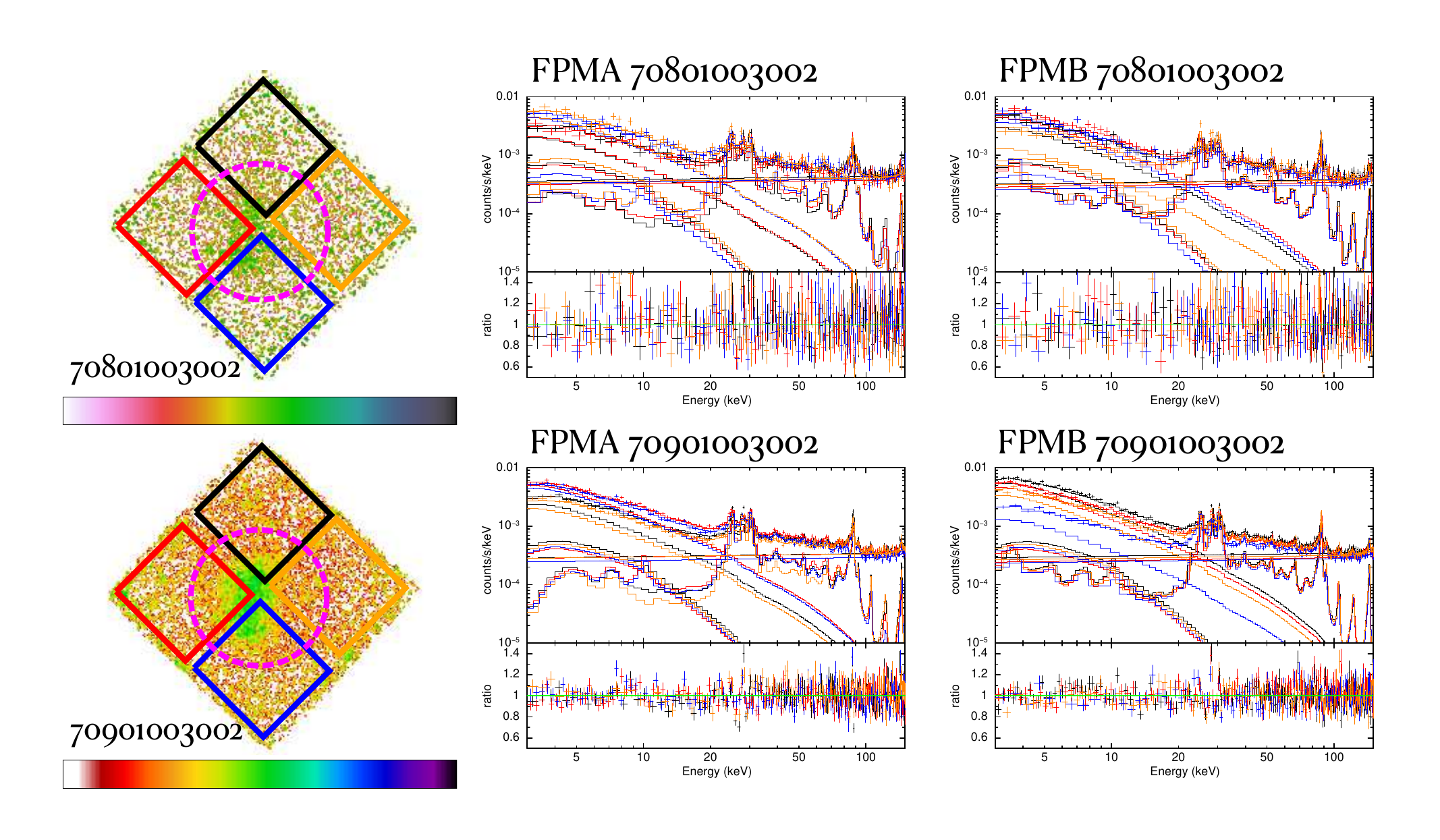}
\caption{Joint-fit of background and cluster emission of \nustar\ FPMA (\textit{upper panel}) and FPMB (\textit{lower panel}). Each color represents a region selected for the background fit that has one-to-one correspondence between the figure and spectra. The left panel indicates the selected background region where the colors correspond to the spectra with the same colors. For plotting purposes, adjacent bins are grouped until they have a significant detection at least as large as 8$\sigma$, with maximum 12 bins. \label{fig:nuskybgd}}
\end{figure*}

We applied \textsc{nuskybgd} to the spectra extracted from each chip of each FPM, aiming to exclude as much emission as possible from the bright central cluster ICM emission as shown in Fig.~\ref{fig:nuskybgd}. We used an {\tt apec} model with free temperature (\textit{kT}) and normalization parameters to account for any residual cluster emission. The joint fit of the background models describing instrumental, aperture and focused X-ray background as well as the contribution from the residual ICM emission for each FPM are shown in Fig.~\ref{fig:nuskybgd}. 

\section{Pre-shock Region Individual Spectra}
\label{sec:app:c}
In this section, we display spectra for FPMs of Region 10 and Region 13 that are displayed together in Fig.~\ref{fig:Reg10n13fit}. These figures also display the background emission.

\begin{figure*}[h!]
\centering
\includegraphics[width=180mm]{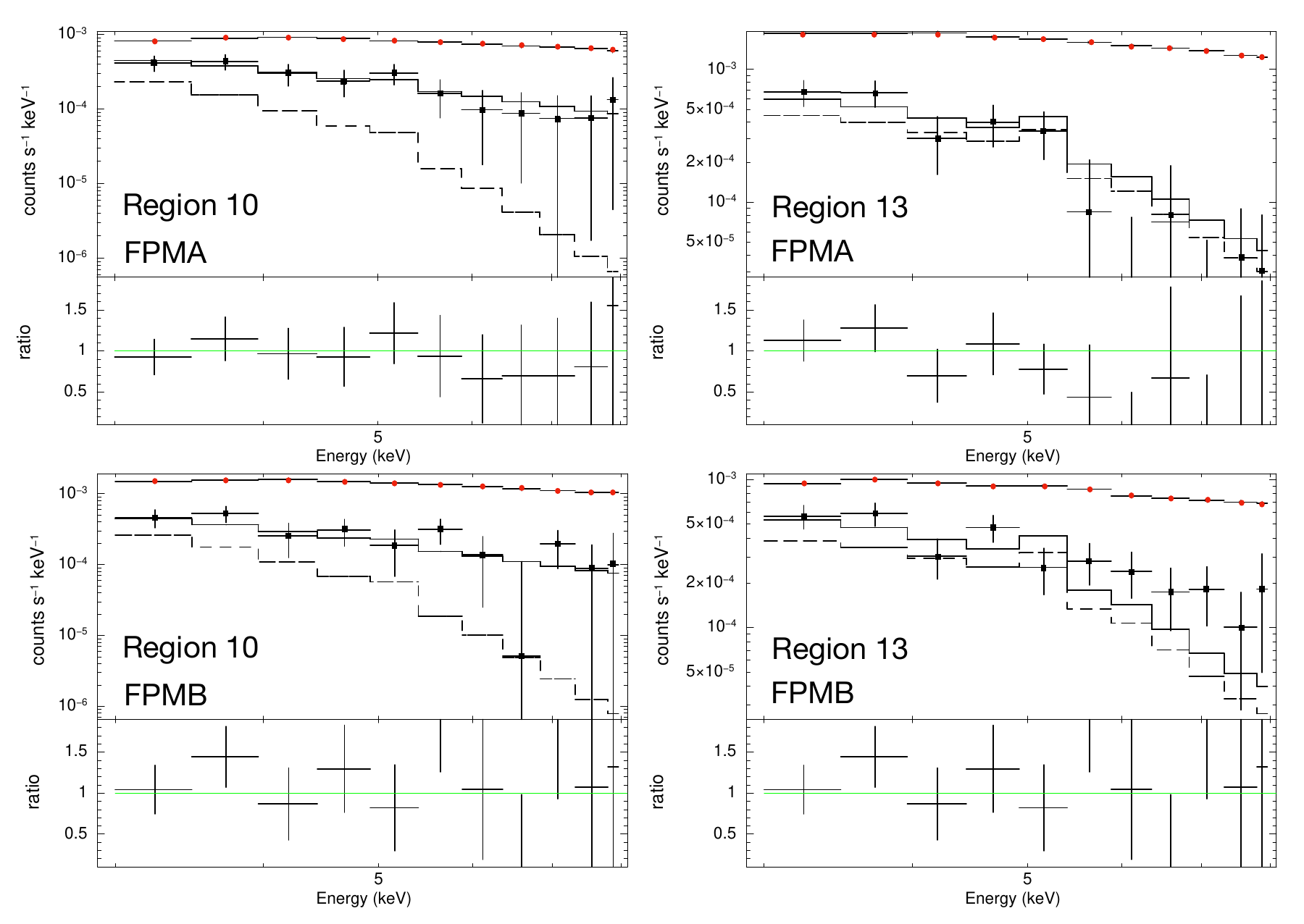}
\caption{The detailed plot of the fit shown in Fig~\ref{fig:Reg10n13fit} including background. Spectral fit for Regions 10 and 13 after background variation and energy band restriction are applied. Black square marks represent the data, solid lines represent the total model, all ARF contributions, and dashed lines the model convolved with only the \textit{intrinsic} ARF; the ARF of the region of interest. The ratio panels represent the ratio of data to the model, and the background model is shown with red circle marker. For plotting purposes, adjacent bins are grouped until they have a significant detection at least as large as 10$\sigma$, with maximum 12 bins. \label{fig:reg10n13back}}
\end{figure*}
\end{document}